# Sub-nanometer depth resolution and single dopant visualization achieved by tilt-coupled multislice electron ptychography

Zehao Dong[1], Yang Zhang[1], Chun-Chien Chiu[2], Sicheng Lu[1], Jianbing Zhang[1], Yu-Chen Liu[2], Suya Liu[3], Jan-Chi Yang[2,4], Pu Yu[1], Yayu Wang[1,5,6], Zhen Chen[7,8]*

[1]State Key Laboratory of Low Dimensional Quantum Physics, Department of Physics, Tsinghua University, Beijing 100084, P. R. China
[2]Department of Physics, National Cheng Kung University, Tainan 70101, Taiwan
[3]Nanoport, ThermoFisher Scientific Shanghai, P. R. China
[4]Center for Quantum Frontiers of Research & Technology (QFort), National Cheng Kung University, Tainan 70101, Taiwan
[5]New Cornerstone Science Laboratory, Frontier Science Center for Quantum Information, Beijing 100084, China
[6]Hefei National Laboratory, Hefei 230088, China
[7]Beijing National Laboratory for Condensed Matter Physics, Institute of Physics, Chinese Academy of Sciences, Beijing 100190, P. R. China
[8]School of Physical Sciences, University of Chinese Academy of Sciences, Beijing 100049, P. R. China

*Corresponding author. Email: zhen.chen@iphy.ac.cn

**Abstract**

Real-space imaging of three-dimensional atomic structures is a critical yet challenging task in materials science. Although scanning transmission electron microscopy has achieved sub-angstrom lateral resolution through techniques like electron ptychography[1,2], depth resolution remains limited to only 2 to 3 nanometers with a single projection setup[3,4]. Attaining better depth resolution typically necessitates large sample tilt angles and many projections, as seen in atomic electron tomography[5,6]. Here, we develop a new algorithm based on multislice electron ptychography which couples only a few projections at small tilt angles, but is sufficient to improve the depth resolution by more than threefold to the sub-nanometer scale, and potentially to the atomic level. This technique maintains high resolving power for both light and heavy atoms, and significantly improves the visibility of single dopants. We are thus able to experimentally detect dilute substitutional praseodymium dopants in a brownmillerite oxide, $Ca_2Co_2O_5$, in three dimensions and observe the accompanying lattice distortion. This technique requires only a moderate level of data acquisition or processing, and can be seamlessly integrated into electron microscopes equipped with conventional components.

**Main Text**

Determining the three-dimensional structure of materials is a critical yet formidable challenge, particularly when endeavoring to visualize the embedded dopants or vacancies within bulk samples, which could be invaluable for research in condensed matter physics[7], chemistry[8], and semiconductor industries[9,10]. Despite the remarkable capabilities of state-of-the-art aberration-corrected scanning transmission electron microscopy (STEM), which has already achieved sub-angstrom resolution in lateral dimensions[11,12], the resolution along beam direction, also known as depth resolution, remains significantly inferior using conventional optical depth-sectioning methods in STEM[13–15]. Another common approach is electron tomography, which has enabled atomic resolution in all three dimensions[5,16], but requires a large number of projections and high tilt angles with only a limited field of view[17]. Recent developments in multislice electron ptychography (MEP), on the other hand, have demonstrated a promising depth resolution of better than 3 nm in thick crystalline samples via only a single projection, together with the ability to resolve light elements such as O, B and N[2,3,18].

The primary factor limiting depth resolution in STEM imaging is the lack of information transfer at high spatial frequencies along the projection direction[3]. This limitation is typically mitigated by increasing the probe-forming semi-angle with advanced aberration correctors[19]. The corresponding depth resolution achieved is around 2.1 nm via optical depth sectioning[4], but this method is only suitable for thin samples due to the limited depth of field and electron channeling effects[20]. Here, we address these issues by introducing an alternative approach called tilt-coupled multislice electron ptychography (TCMEP), in which a moderate probe-forming semi-angle is used while the sample is intentionally tilted off-axis to capture information from higher angles. Simulations indicate that the improvement is most prominent for small tilt angles, with atomic-scale depth resolution achievable at tilt angles of 4°. Our experiments with TCMEP successfully achieve sub-nanometer depth resolution, and effectively transfer information from higher frequencies along the projection through the sample tilt series. This enhancement significantly improves the identification of single dopants and the associated lattice distortions in three dimensions, providing valuable insights into the physical properties of materials. TCMEP requires only small tilt angles (~4°) and few scans (fewer than 5 scans in this work), making it compatible with widely available STEM instruments equipped with conventional double-tilt sample holders.

**Principle and reconstruction process**

In TCMEP, we simultaneously reconstruct a shared sample object function using a few four-dimensional STEM (4D-STEM) datasets. These datasets are acquired within the same sample region projected along axes intentionally tilted away from the zone axis by a small angle much less than 1 radian (Fig. 1a). The residual relative shift among datasets is corrected using a pre-reconstruction alignment procedure, followed by further refinements during the ptychographic reconstruction (details in Methods). A similar approach in X-ray ptychography has demonstrated efficient filling of the missing cone in 3D information transfer using this strategy[21]. In the following, we will show that the implementation of sample tilts in electron ptychography opens up new imaging possibilities, in particular the three-dimensional resolving power at a mild illumination, which, however, has never been explored.

To illustrate the advantages of TCMEP, we shift to the Fourier space to analyze the scope of information obtained from the reconstructed result (Fig. 1b). The 3D information transfer via MEP from a single dataset is conceptually represented as a wedge, delineated by an effective semi-angle $α_{eff}$. This information boundary surpasses the $2α$ contrast transfer in optical sectioning methods within the linear imaging approximation[2,22,23]. When the sample experiences a slight tilt by an angle $θ$ in real space, the corresponding wedge in Fourier space undergoes a tilt of the same angle. As depicted in the middle and bottom panels of Fig. 1b, coupling datasets with tilt angles $θ$ and $-θ$ results in the fusion of their information, generating an expanded wedge of information transfer with an increased semi-angle of around $α_{eff} + θ$. This process effectively captures features with higher spatial frequencies along the $z$-axis, thereby enhancing depth resolution.

The reconstruction process of TCMEP is displayed in the flow chart shown in Fig. 1c. In contrast to conventional MEP, our new algorithm operates in a parallel iterative manner, in which $N$ (scan number) 4D-STEM datasets captured under various tilt conditions are coupled to reconstruct a unified multislice object. Due to the relatively small tilt angle ($\ll 1$ rad), we introduce an interlayer shift as an approximation to generate a tilted model of a shared object. Subsequently, within each iteration, $N$ different probes for each scan propagate in parallel through these $N$ tilted objects, while the exit waves are utilized to calculate the loss function and determine the update of object and probes. The iteration continues until a convergent result is achieved. The phase of the resulting complex object function represents the atomic potential in the sample and reflects the distribution of atomic defects[2,24], which is utilized in our analyses below.

**Simulation on imaging a single dopant**

We begin by evaluating the performance of TCMEP through simulations on a $SrTiO_3$ crystal with artificially introduced interstitial and substitutional dopants. In Fig. 2a-c, we compare phase images at the same depth under three distinct conditions: maximum tilt angles of 0°, 2°, and 4°. The total electron dose ($2.5×10^6$ $e/Å^2$) remains constant for all three cases. All atomic columns, including oxygen, are resolved, with an additional weak feature (highlighted with circles) corresponding to a single Sr dopant incorporated into the structural model. Notably, with increasing tilt angle, the visibility of the dopant atom is significantly enhanced, primarily due to improved depth resolution which reduces the depth-wise blurring of the single atom. This is illustrated in Fig. 2d through the phase-depth curve of the dopant atom, which exhibits the characteristic profile of a resolution-blurred Gaussian function and sharpens at higher tilt angles. The enhancement in visibility is also observed for dopants or vacancies of lighter elements such as O (Fig. S1), indicating the augmented depth resolution achieved by TCMEP.

To show the advantages of TCMEP in dose efficiency, we also carried out simulations under various illumination dose conditions. Figure 2e presents the tabulated depth sectioning images of the Sr dopant at total electron doses of $2.5×10^4$ $e/Å^2$, $2.5×10^6$ $e/Å^2$, and $2.5×10^8$ $e/Å^2$. At the low electron dose level of $2.5×10^4$ $e/Å^2$, the dopant is barely distinguishable with conventional MEP. Notably, with a slight tilt of 2°, it becomes discernible, and its contrast further intensifies at a maximum tilt angle of 4°. The broadening of the dopant atom along the depth dimension is also significantly reduced. The images taken at $2.5×10^6$ $e/Å^2$ and $2.5×10^8$ $e/Å^2$ under the same tilt conditions are nearly identical, with remarkable improvements at higher tilt angles in both cases. Therefore, the typical experimental dose ($\sim 10^6$ $e/Å^2$) is sufficient for obtaining atomic level depth resolution ($\sim$ 4 Å) with TCMEP using only 5 tilts up to 4° in our simulations. It is noted that dopant atoms are hardly distinguishable from the artifacts at an illumination dose of around $10^3$ $e/Å^2$ using present imaging conditions.

To analyze the simulation results on a quantitative level, we fit all the phase-depth curves with a Gaussian function $y = A \exp\left[-\frac{(x-\mu)^2}{2\sigma^2}\right] + B$, and determine the depth resolution ($d=1.33\sigma$) using the convention from a previous work[2]. As illustrated in Fig. 2f, with a sufficiently large electron dose and a maximum tilt angle of 4°, $d$ is improved by a factor of 2 to 3, ultimately achieving atomic-scale resolution of approximately 0.36 nm. On the low dose side with comparable tilt angles, the improvement of TCMEP is more significant, reaching a depth resolution of around 0.55 nm, which is more than 4 times better than that achieved with MEP.

**Depth resolution to sub-nanometers**

Next, we move on to experimental results on a twisted bilayer $SrTiO_3$ sample with a relative twisting angle of approximately 9°. Twisted bilayer samples exhibit exotic quantum many-body phenomena and have potential applications in twistronics[25–27]. Notably, the spatial variation of the buried interface can only be efficiently probed in STEM with depth-resolving techniques[18]. The fabricated sample features a clean and sharp interface, with each layer a few nanometers thick. In the projection (Fig. 3i), a Moiré pattern emerges due to the intermixing of the top and bottom layers, whereas ideal depth sectioning could clearly separate the top and bottom layer without forming Moiré patterns. Therefore, this system can be used to test depth resolution by monitoring the residual layer intermixing near the interface. The improvement in depth resolution is evident in real space images shown in Fig. 3a-h, in which four layers (each with a thickness of 6.4 Å) near the interface are displayed from two distinct reconstruction results with maximum tilt angles 0° and 2° (full reconstruction shown in Fig. S2-3). The Moiré pattern is scarcely discernible in the first slice (Fig. 3a and e), but becomes notably apparent in the second slice (Fig. 3b and f). With a maximum tilt angle of 2°, the Moiré pattern diminishes rapidly and ultimately vanishes entirely in the last slice (Fig. 3h). On the contrary, in the untilted reconstruction, a faint residual Moiré pattern remains in the last slice (Fig. 3d). This indicates an enhanced separation between the top and bottom layers of the sample with TCMEP. The discrepancy in average phase-depth curves further supports the improvement in depth resolution at higher tilt angles (Fig. 3j). By fitting the curve associated with a 2° tilt angle, we achieve an unprecedented depth resolution of ~0.9 nm at the interface with a total imaging dose of $8.4\times10^5$ $e/Å^2$ (details in Fig. S4). Notably, the depth resolution surpasses the chromatic aberration limit of ~1.4 nm, and agrees well with our simulation under comparable conditions.

The Fourier transform of the reconstructed phase image captures information in the reciprocal space, which is subsequently utilized to extract the 'missing wedge' as displayed in Fig. 3k (details in Fig. S5). The results qualitatively agree with the schematic illustration shown in Fig. 1b, in which the information boundary of each reconstruction exhibits a wedge-shaped pattern. With increased tilt angles, the depth resolution improves across all lateral spatial frequencies. Fitted by a linear function (as denoted by the dashed line), the effective semi-angle of the information boundary, $\alpha_{eff}$, is inferred from the slope of each curve (inset of Fig. 3k). It is observed that $\alpha_{eff}$ increases by approximately the same amount whenever the maximum tilt angle increases. This provides compelling evidence that the improvement in depth resolution predominantly stems from information gathered at larger angles.

**Imaging dopant atoms and 3D lattice distortions**

TCMEP is then applied to image dopant atoms, which can strongly modify the emergent phenomena of quantum materials[7]. Previous reports on cobalt oxides in the brownmillerite phase revealed a tunable magnetic ground state upon doping[28]. Our experiments focus on a specific brownmillerite $(Pr_xCa_{1-x})_2Co_2O_5$ thin film (nominal $x$ is about 0.05) grown on a $LaAlO_3$ (001) substrate (Fig. 4a), which consists of alternatively-stacked $CoO_6$ octahedra and $CoO_4$ tetrahedra[29,30]. We used only three tilt angles to minimize the experimental workload (details in Fig. S6). Figure 4b displays the summed phase image reconstructed with TCMEP, which exhibits sufficient lateral resolution with all atomic columns differentiated. Two representative phase-depth curves of Ca columns are displayed in Fig. 4c. The blue curve exhibits a standard pattern with a plateau in the sample area, whereas the red curve reveals an additional peak in the depth range between 18 nm and 20 nm, which is attributed to the substitution of Pr for Ca atoms.

To exclude possible experimental or numerical artifacts, we perform the same analysis on the Co columns, as shown in Fig. 4d (full datasets in Fig. S7-8). In all Co columns within our reconstruction, the plateau remains consistently flat and well-defined, in stark contrast to the Ca columns where the presence of an additional peak is quite common. Furthermore, the phase histogram for all Ca atoms (Fig. 4e) exhibits a distinct shoulder adjacent to the primary peak, which is well-fitted by a secondary Gaussian component. The Pr doping level is subsequently estimated as 8.8%±2.9%, roughly in agreement with the nominal concentration of 5%. In contrast, the histogram for Co columns (Fig. 4f) is symmetric and lacks additional features. Once again, it is crucial to emphasize the importance of TCMEP in detecting Pr dopants: the histogram of Ca columns from the normal MEP reconstruction (inset of Fig. 4e) reveals minimal asymmetric features without statistical significance.

The summed phase image along the depth dimension is displayed in Fig. 5a, with a modified colormap to enhance contrast. Several Ca rows exhibit higher phase values (e.g., the lower Ca row selected by the white dashed rectangle), indicating a significant presence of Pr dopants. To visualize the distribution of Pr dopants along the depth dimension, we present depth sectioning along the two selected Ca rows in Fig. 5b-c. The upper Ca row shows minimal phase fluctuation, while the lower row displays a characteristic pattern of Pr dopants distributed randomly along the depth dimension. Phase-depth curves for three selected Ca columns are plotted in Fig. 5f, with Pr dopants marked by arrows. Notably, TCMEP circumvents the channeling effect[2], enabling the identification of multiple dopant atoms within the same atomic column.

As a result, all Pr dopants are unambiguously identified in three dimensions by setting a threshold phase based on the statistics in Fig. 4e. The accompanying depth-dependent atomic displacements are also extracted. Two representative slice images are displayed in Fig. 5d-e, with Pr dopants and atomic displacements highlighted with red circles and arrows, respectively (full dataset in Supplementary Movie S1). The characteristic lattice distortions, on the order of 10 pm, are likely induced by the Pr dopants. Two schemes may be relevant for the distortions upon doping: (i) the modified Co-O bond lengths according to bond valence sum approaches[31]; (ii) the formation of Jahn-Teller (JT) active $Co^{2+}$ ions ($3d^7$ configurations), which lift the orbital degeneracies through JT distortions[32].

**Discussion**

We have demonstrated that TCMEP significantly improves depth resolution and the visibility of single dopants. This ability to image the distribution of atomic defects paves the way for understanding various physical properties of materials, from semiconductor devices[9] to high-temperature superconductors[33]. The improvement in depth resolution to sub-nanometer scales also enables new capabilities to resolve complex 3D structures, such as nitrogen-vacancy centers[34], topological polar textures[35], and nanoscale phase separations[36].

Although TCMEP could be limited by its dependence on a sufficiently small tilt angle for the validity of the interlayer shift approximation, reconstruction with a maximum tilt angle of 10° (~0.17 rad, which can still be regarded as a small value) is achievable and reliable, with a depth resolution better than 3 Å, as suggested by the simulations in Fig. S9. Fortunately, TCMEP can be extended to even larger tilt angles with the implementation of projection algorithms used in tomography[37–39]. Several experimental results based on sequential rather than joint reconstructions combining MEP and tomography have been reported, showcasing only a moderate resolution of 1.75 Å and precision of 17 pm in three dimensions[6,40]. Light atoms like oxygen remained unresolved in these previous reports, likely due to uncertainties in the registration of scan positions and sample tilts. TCMEP, on the other hand, significantly reduces the complexity of data acquisition and processing, as well as the overall workload. Despite a slight sacrifice in depth resolution, the results demonstrate much better overall quality and precision.

In summary, we have introduced sample tilt-series to multislice electron ptychography, which gathers information from higher angles and improves the experimental depth resolution into a few angstroms. This new method performs remarkably well in the imaging of single dopants and the atomic displacements in all three dimensions. It is also dose-efficient and requires only a few tilts, thus compatible with conventional aberration-corrected STEM and double-tilt sample holders routinely available in most laboratories. Further improvements may lead to the attainment of three-dimensional atomic resolution in the future.

# Methods

**Sample growth and preparation for TEM measurements**

Brownmillerite $(Pr_{0.05}Ca_{0.95})_2Co_2O_5$ thin films were grown by a customized pulsed laser deposition (PLD) system, at 640 °C with an oxygen pressure of 0.06 mbar. The laser (KrF, $\lambda = 248$ nm) energy density was set at 1.1 J/cm$^2$ with the repetition rate of 5 Hz. After the growth, the samples were cooled down to room temperature at a cooling rate of 10 °C/min at 0.06 mbar oxygen pressure. The crystalline structures of thin films were characterized by a high-resolution four-circle X-ray diffractometer (Smartlab, Rigaku) using monochromatic Cu $K_{\alpha1}$ radiation ($\lambda = 1.5406$ Å). TEM samples were prepared using a focused ion beam (FIB) instrument (Zeiss Auriga). The samples were thinned down to 100 nm using an accelerating voltage of 30 kV with a decreasing current from 240 pA to 50 pA, followed by a fine polish with an accelerating voltage of 5 kV with a current of 20 pA.

The freestanding SrTiO$_3$ (STO) films were grown by PLD, using KrF (248 nm) excimer laser. The STO thin films were deposited on (La,Sr)MnO$_3$ (LSMO) buffered (001)-oriented STO substrates, carried out under an oxygen pressure of 100 mTorr at a temperature of 700 °C, utilizing a laser power of 250 mJ and a laser repetition rate of 10 Hz. The heterostructure was then immersed in hydrochloric acid to dissolve LSMO sacrificial layer and to separate STO film from the single crystal substrate. The freestanding STO was then transferred onto another single-crystalline STO heterostructure grown on LSMO buffered (001)-oriented STO substrate, having a twist angle with respect to the STO single-crystal substrate. Thereafter, the sample was immersed in hydrochloric acid once again to obtain the freestanding twisted bilayer STO membrane. The STO membrane was then transferred onto a copper TEM grid. Sufficient rinsing was performed to make the interface of the two layers clean without any residual, as illustrated in the ptychographic reconstruction in Fig. 3 and Figs. S2-3.

**Experimental details**

Experiments on the twisted SrTiO$_3$ bilayers were performed on an aberration-corrected Spectra 300 (Thermo Fisher Scientific) electron microscope operating at 300 keV, equipped with an ultrahigh brightness Cold Field Emission Gun (X-CFEG). The experiments on the $(Pr_{0.05}Ca_{0.95})_2Co_2O_5$ thin film were carried out on a Titan Cubed Themis 60-300 (Thermo Fisher Scientific) electron microscope operating at 300 keV with a high-brightness Schottky Field Emission Gun. Throughout our experiments, the probe-forming semi-angle was set to 25 mrad, and the four-dimensional STEM (4D-STEM) datasets were acquired with an Electron Microscope Pixel Array Detector (EMPAD) with 124×124 pixels. The focal point of the probe was set ~20 nm above the sample's top surface. For each 4D-STEM dataset, the total scanning area was 9.3×9.3 nm$^2$, with 200×200 scanning points distributed uniformly, and the dwell time for each diffraction pattern was 1 ms. The beam current was 30 pA for the $(Pr_{0.05}Ca_{0.95})_2Co_2O_5$ film (corresponding dose is $9.0\times10^5$ $e$/Å$^2$), but was reduced to 7 pA for the SrTiO$_3$ sample to minimize beam damage (corresponding dose was $2.1\times10^5$ $e$/Å$^2$).

For experiments on twisted bilayer SrTiO$_3$, tilt angles of ±2°, ±1° and 0° were acquired, and we respectively utilized 1 scan (0°), 2 scans (±1°), and 4 scans (±1°, ±2°) for MEP or TCMEP reconstructions. For experiments on $(Pr_{0.05}Ca_{0.95})_2Co_2O_5$, tilt angles of ±1° and 0° (from the [100] zone axis) were acquired, and we respectively utilized 1 scan (0°) and 3 scans (±1°, 0°) for MEP or TCMEP reconstructions. The depth resolution was around 2.6 nm for MEP and 1.5 nm for TCMEP (Fig. S7). The enhanced depth resolution surpasses the improvement from merely raising the illumination dose (as the simulation in Fig. 2f suggests). In our experiments, accurate tilt angles were not required, as further refinements were performed in the reconstruction algorithm.

**Alignment of different tilts and the reconstruction process**

Before acquiring the 4D-STEM data, we focused our probe on a fixed spot on the sample near the region of interest (ROI) for about 15 minutes. As shown in Fig. S7a, a distinct structural feature (a hole with a dislocation) was formed, which was later used to track the ROI after tilting the sample. A precise alignment among these datasets was achieved by further refinements before the TCMEP reconstruction. First of all, a conventional MEP reconstruction was performed on each dataset and used to determine the exact position of the reference point. Then, all of the datasets were aligned accordingly and cropped into a smaller overlapped region of 100×100 scanning points. The TCMEP reconstruction was performed on those well-aligned and overlapped datasets. The precise scanning positions from different datasets were further refined during the TCMEP reconstruction via a drift-correction algorithm. The probe corresponding to each scan was modeled with four modes under the mixed-state algorithm[41]. The update direction and step within each iteration was evaluated with the least-squares maximum likelihood (LSQML) method[42–44].

Notably, The TCMEP reconstruction in this work is time-intensive and requires large GPU memory. All numerical processing was performed with an Nvidia A100 GPU with 80 GB of memory. Careful selection of computational parameters is essential to avoid memory overflow issues, given the exceptionally large datasets—each approximately 2 GB—and the extensive number of slices (up to 80, at 2 Å per slice) necessary to achieve superior depth resolution. Additionally, the reconstruction procedure is time-consuming. For instance, it takes roughly 30 hours with 1000 iterations to obtain the final result with a maximum tilt angle of 2° as shown in Fig. 3. Therefore, advancements in both hardware capabilities and algorithmic efficiency are crucial to accelerate the computational process for larger datasets, especially when striving for the ultimate atomic-scale depth resolution.

**4D-STEM Simulation**

The simulated 4D-STEM datasets were all generated at 300 keV beam energy using the μSTEM software[45]. The probe's convergent semi-angle was 25 mrad, and was overfocused by 20 nm above the sample surface. 26×26 diffraction patterns with a 0.60 Å step were generated at different sample tilt angles. We used a 12.5 nm-thick $SrTiO_3$ structural model along the [001] zone axis with different artificially introduced dopants. The structural model contained 4×4 unit cells in real space (15.6×15.6 Å$^2$ in area), and was sampled with 128×128 pixels. The corresponding reciprocal space sampling was 0.0641 Å$^{-1}$ per pixel, and the number of pixels was the same as in real space. The impact of illumination dose was simulated by incorporating Poisson noise into the datasets accordingly. For each TCMEP simulation with multiple datasets, the interval of tilt angle was taken as 2°. The slice thicknesses chosen in reconstructions were 3.6 Å (for single 0°), 2.7 Å (for a maximum angle of 2°), and 1.8 Å (for a maximum angle of 4° data). The slice thicknesses were smaller than the Nyquist sampling rate for depth resolution, therefore would not modify the reconstructed results.

**Precision in measuring 3D atomic positions**

The precision in measuring 3D position of atoms is tolerable with TCMEP, with an accuracy of 1.8 pm in-plane and 0.3 nm in depth, estimated from the peak positions of Sr/Ti atoms (Fig. S10). These values are both comparable to MEP results. Notably, the in-plane precision of TCMEP (1.8 pm) is slightly worse than that of MEP (1.3 pm), which can be partly attributed to the slight misalignment among datasets and the extra process of position correction. It should also be noted that the estimation for precision in the depth dimension is only an upper limit, due to the inevitable surface roughness and local curvature of the interface beyond the measurement uncertainty. Moreover, the statistical variation of 0.3 nm in depth indicates that the surface roughness of the fabricated $SrTiO_3$ sample is smaller than the size of a single unit cell (0.4 nm).

## Data and materials availability

The code and raw data presented in this study are available upon reasonable request from the corresponding authors.

## Acknowledgments

This work was supported by the National Key Research and Development Program of China (MOST) (Grant No. 2022YFA1405100), the National Natural Science Foundation of China (Grant No. U22A6005, No. 52273227 and No. 52025024), the Basic Science Center Project of NSFC (No. 52388201), the Innovation Program for Quantum Science and Technology (No. 2021ZD0302502), and the Basic and Applied Basic Research Major Programme of Guangdong Province, China (Grant No. 2021B0301030003). Y.W. is partially supported by the New Cornerstone Science Foundation through the New Cornerstone Investigator Program and the XPLORER PRIZE. J.-C.Y. acknowledges the financial support from the National Science and Technology Council (NSTC) in Taiwan under grant no. NSTC-112-2112-M-006-020-MY3. This work used the facilities of the National Center for Electron Microscopy in Beijing at Tsinghua University and Nanoport at ThermoFisher Scientific Shanghai.

## Author contributions

Z.C. initiated the idea and designed the studies. Z.D. developed the algorithm, conducted simulations, and analyzed data with the supervision of Z.C. and Y.W.. Electron microscopy experiments were performed by Z.D., Y.Z., S.Y.L., and Z.C.. Pr-doped $Ca_2Co_2O_5$ films were grown by S.C.L., J. Z., and P.Y.. Twisted bilayer STO samples were prepared by C.-C.C., Y.-C.L., and J.-C.Y.. Z.D. and Z.C. wrote the manuscript with inputs from all authors.

## Competing interests

Authors declare that they have no competing interests.

## Additional information

**Supplementary information** is available.

**Correspondence and requests for materials** should be addressed to the corresponding authors.

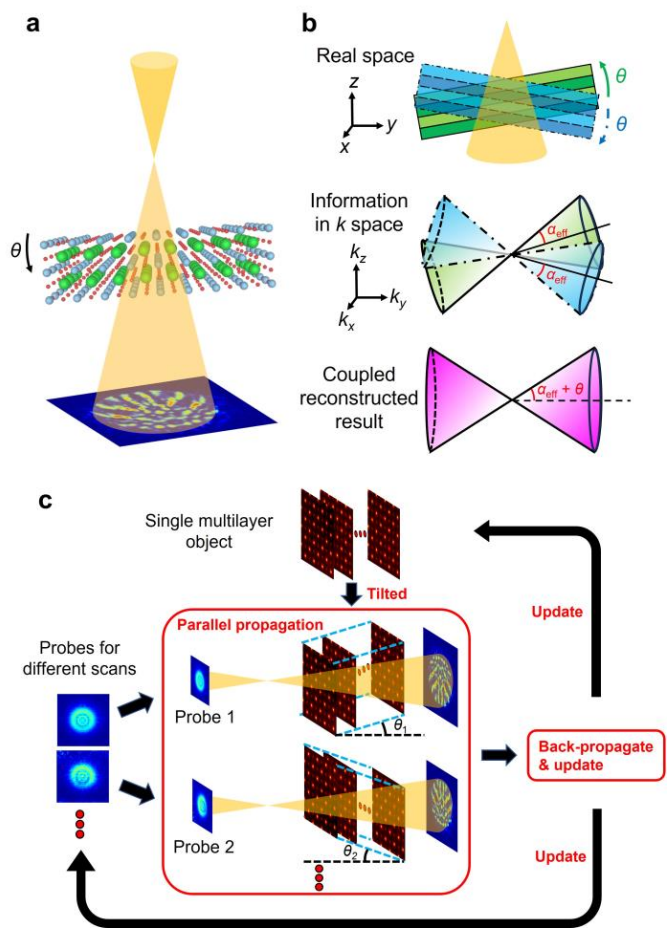

**Fig. 1| Principle and process of tilt-coupled multislice electron ptychography (TCMEP). a**, Experimental setup for TCMEP, where the sample is tilted off the zone-axis by a small angle denoted by $\theta$. Multiple datasets under various tilt conditions are acquired in experiments. **b**, Illustration of the principle on improving the depth resolution with TCMEP. The upper panel displays the sample-tilts with angle $\theta$ relative to the zone axis. The middle panel illustrates the corresponding information transfer under these tilt conditions in Fourier space, which is cone-shaped with an effective semi-angle $\alpha_{eff}$. The lower panel demonstrates the improved information transfer from coupling these datasets, in which the effective semi-angle is enhanced to $\alpha_{eff}+\theta$ through sample-tilts. **c**, Flow chart of the reconstruction algorithm for TCMEP, which is based on a parallel reconstruction using all datasets for optimizing a single multilayer object.

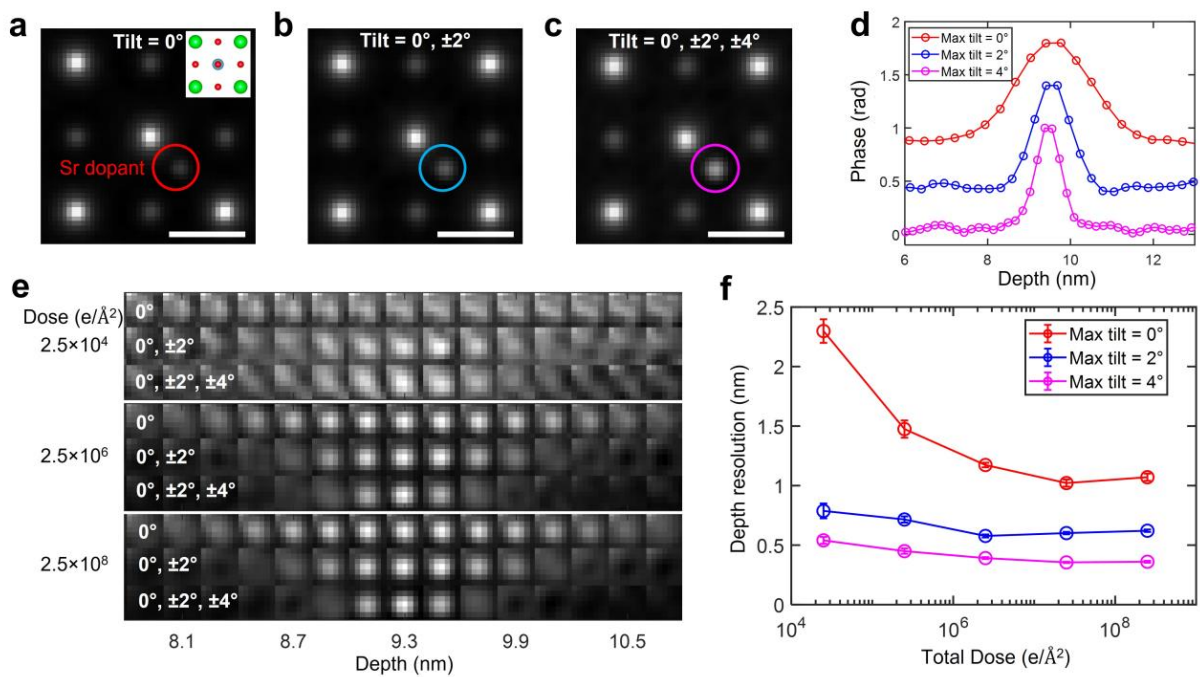

**Fig. 2| Simulation on imaging single dopants with TCMEP. a-c**, Reconstructed phase images of the same slice containing a Sr dopant (marked with circles), with tilt angle 0° (a), 0° and ±2° (b), 0°, ±2°, and ±4° (c). Total dose is maintained as $2.5\times10^6$ $e$/Å$^2$. The inset of (a) displays the corresponding atomic structure. Scale bars, 2 Å. **d**, Phase-depth plot of the Sr dopant under the above three conditions. Curves are vertically shifted for clarity. **e**, Depth section of an interstitial Sr atom along the horizontal axis with a 2.1 Å step (slice thickness). Results collected at different tilt angles and electron doses are stacked along the vertical axis. **f**, Fitted depth resolution under different conditions of simulations. Error bars are determined from curve fitting.

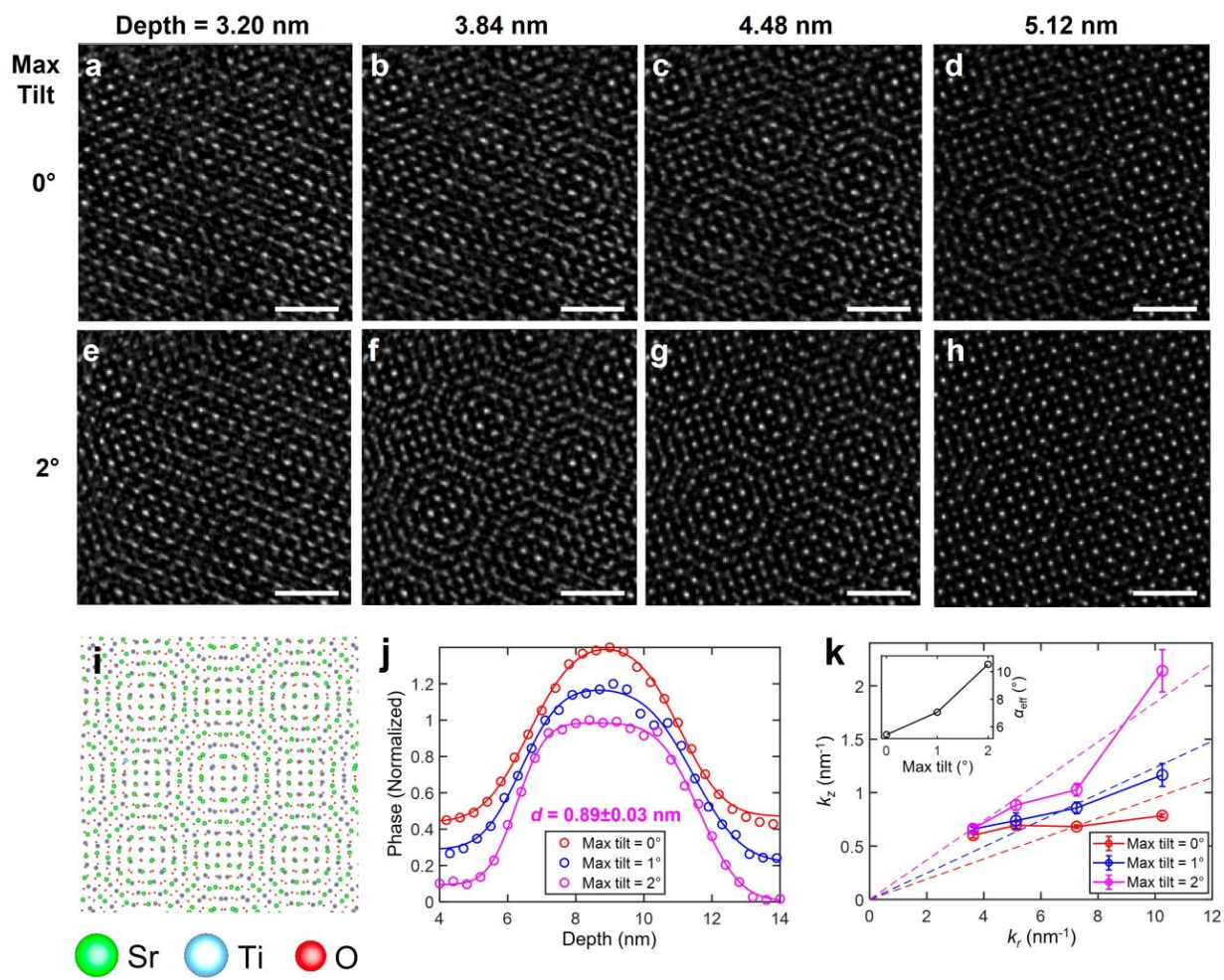

**Fig. 3| Experimental demonstration of missing wedge filling and improvement of depth resolution. a-h**, Reconstructed phase images adjacent to the Moiré interface, with depth and tilt conditions annotated for each column and row. Scale bars, 1 nm. **i**, Crystal structure of the twisted bilayer $SrTiO_3$. **j**, Averaged phase-depth curves from Sr and Ti atoms in the bottom layer, fitted with error functions. A depth resolution of ~0.9 nm is found on the interface with TCMEP under a maximum tilt angle of 2°. **k**, Extracted boundary of information transfer for TCMEP reconstruction with a maximum tilt angle 0°, 1° and 2°, respectively. Dashed lines are linear fits for the curves. The inset shows the effective semi-angle of information transfer versus the maximum tilt angle. Error bars are determined from curve fitting (details shown in Fig. S5).

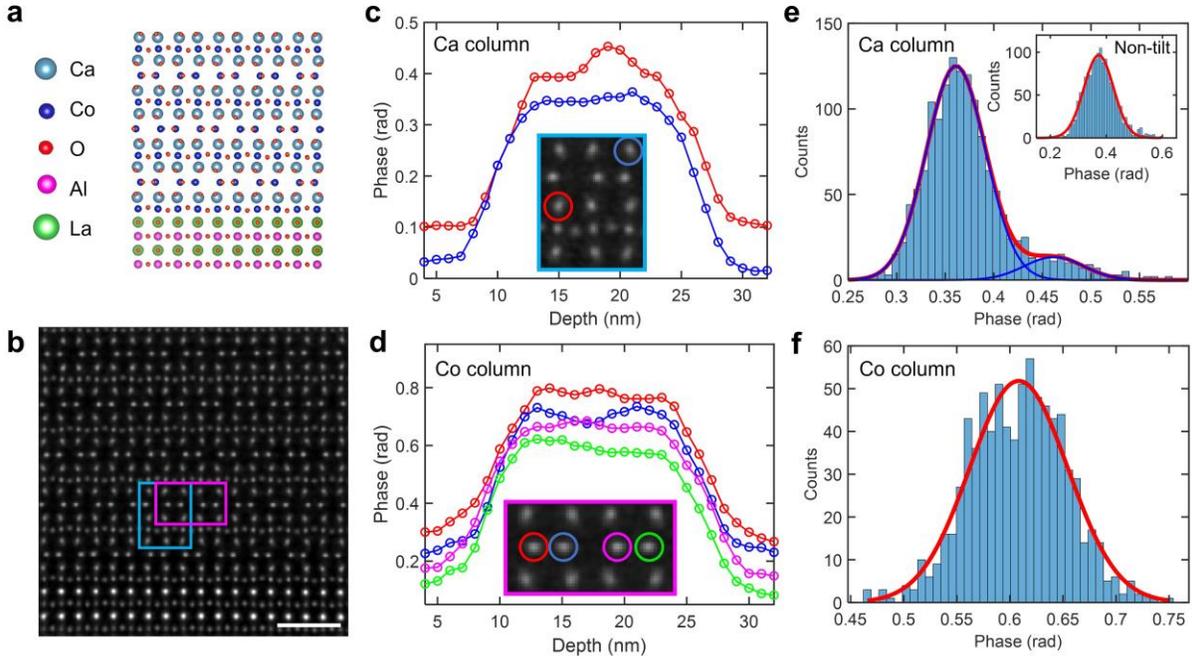

**Fig. 4| Experimental imaging of the Pr dopants with TCMEP. a**, Crystal structure of the $Ca_2Co_2O_5$ film grown on a $LaAlO_3$ substrate. **b**, Summed phase image reconstructed by TCMEP with tilt angles 0° and ±1°. Scale bar, 1 nm. **c-d**, Phase depth curves displaying characteristic Ca (**c**) and Co (**d**) columns. The insets are magnified view of the regions selected in (**b**). Each curve corresponds to the atom selected by the circle of same color, which is vertically shifted for clarity. **e-f**, Histograms of the phases of Ca (e) and Co (f) atoms from slice images between $z = 15$ nm and $z = 23$ nm. The inset of (**e**) shows the histogram of Ca atoms from conventional MEP. All histograms are fitted with either one or two Gaussian components, as indicated by the red and blue curves.

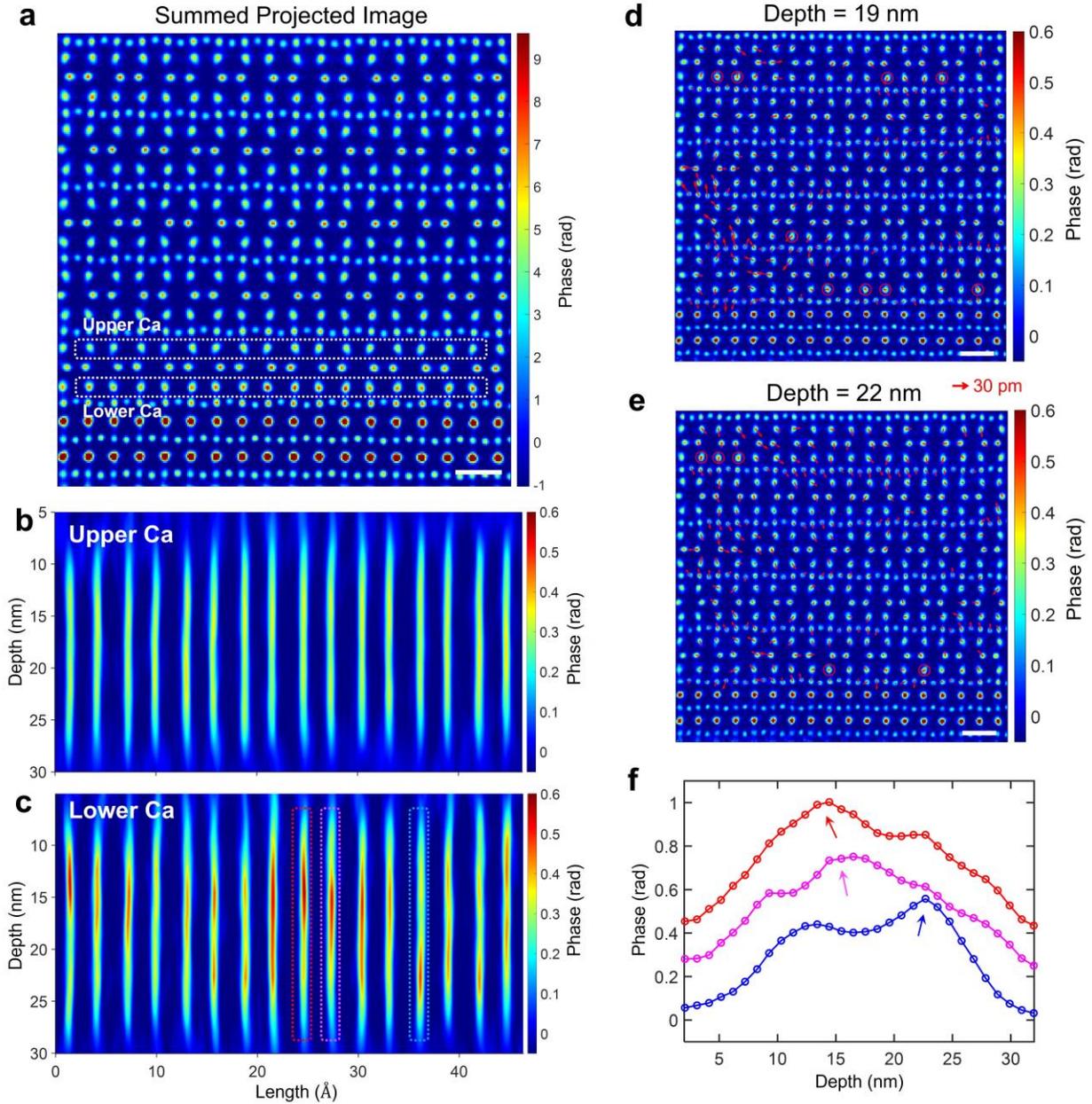

**Fig. 5| Distribution of dopants and the accompanying lattice distortions. a**, Image of total phase projected along the depth dimension. Two rows of inequivalent Ca columns are selected by the white dashed rectangle. Scale bars, 5 Å. **b-c**, Depth sectioning plot of the two Ca rows selected in (a). The upper Ca row (b) exhibits a smaller fluctuation in phase than the lower one (c). **d-e**, Slice images from depths of 19 nm (d) and 22 nm (e), respectively, with a slice thickness of 1 nm. Red circles denote the positions of Pr dopants, and red arrows indicate the atomic displacement vectors. Scale bars, 5 Å. **f**, Depth section curves for the three columns selected by the corresponding rectangles in (c), exhibiting a Pr substitution dopant at different depths.

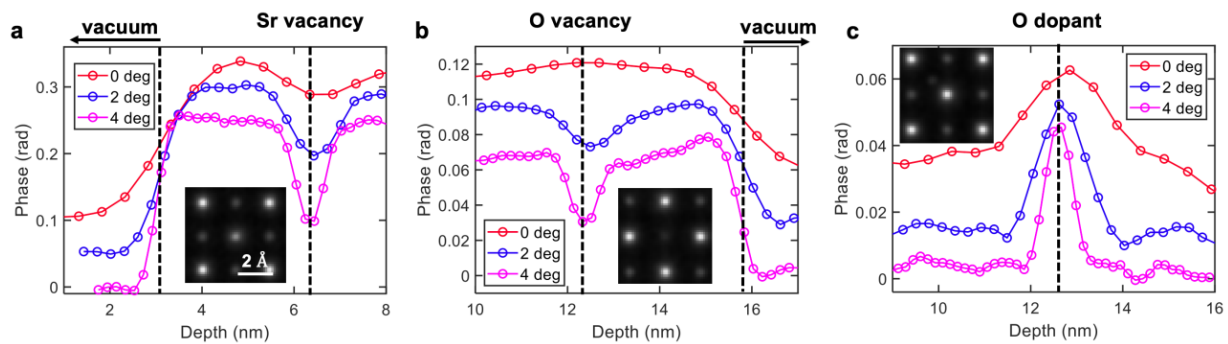

**Fig. S1| Simulation on imaging various types of atomic defects. a**, Phase-depth curves of the Sr column with a vacancy, reconstructed using maximum tilt angles of 0°, 2°, 4°. The upper surface of the sample and the position of Sr vacancy are marked with dashed lines. The inset shows the phase image of the slice from the depth of Sr vacancy. **b-c**, The same as (a), but for O vacancies and O dopants.

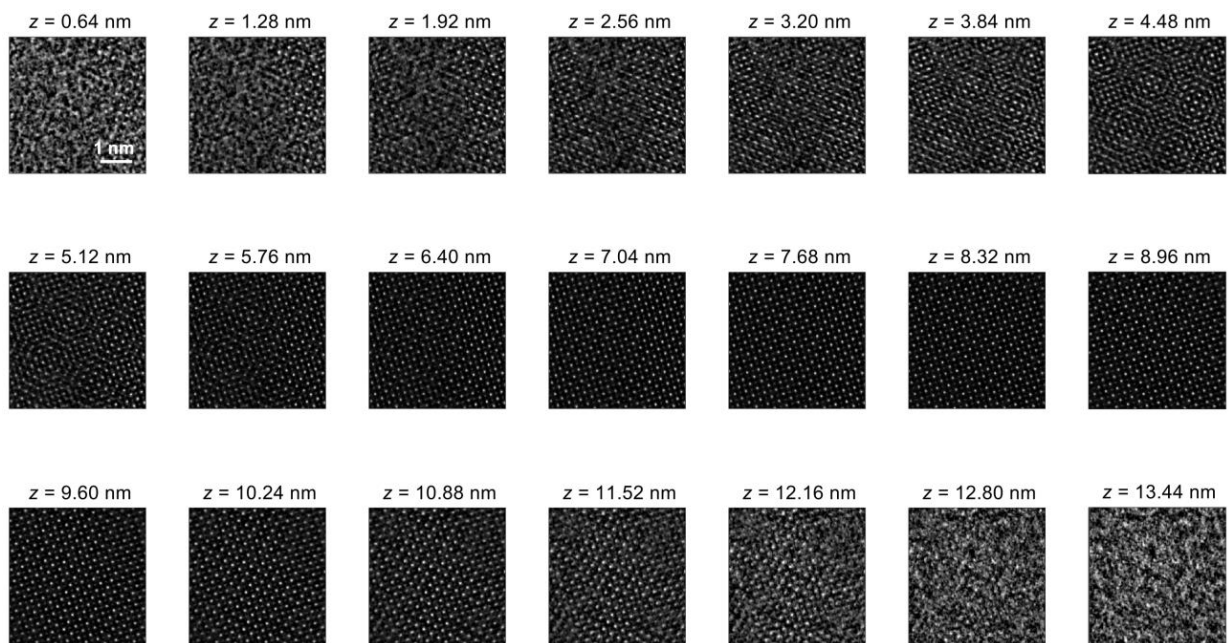

**Fig. S2| Full reconstructed dataset for Fig. 3a-d in the main text.** The maximum tilt angle is 0°, and the corresponding depth is annotated above each slice.

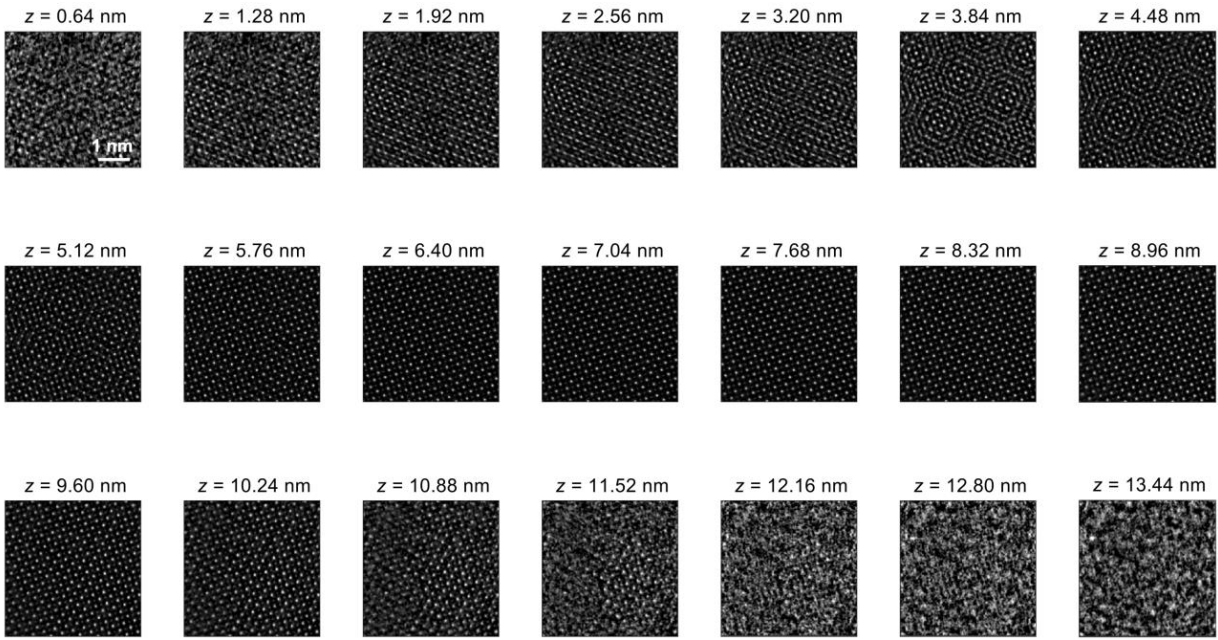

**Fig. S3| Full reconstructed dataset for Fig. 3e-h in the main text.** The maximum tilt angle is 2°, and the corresponding depth is annotated above each slice.

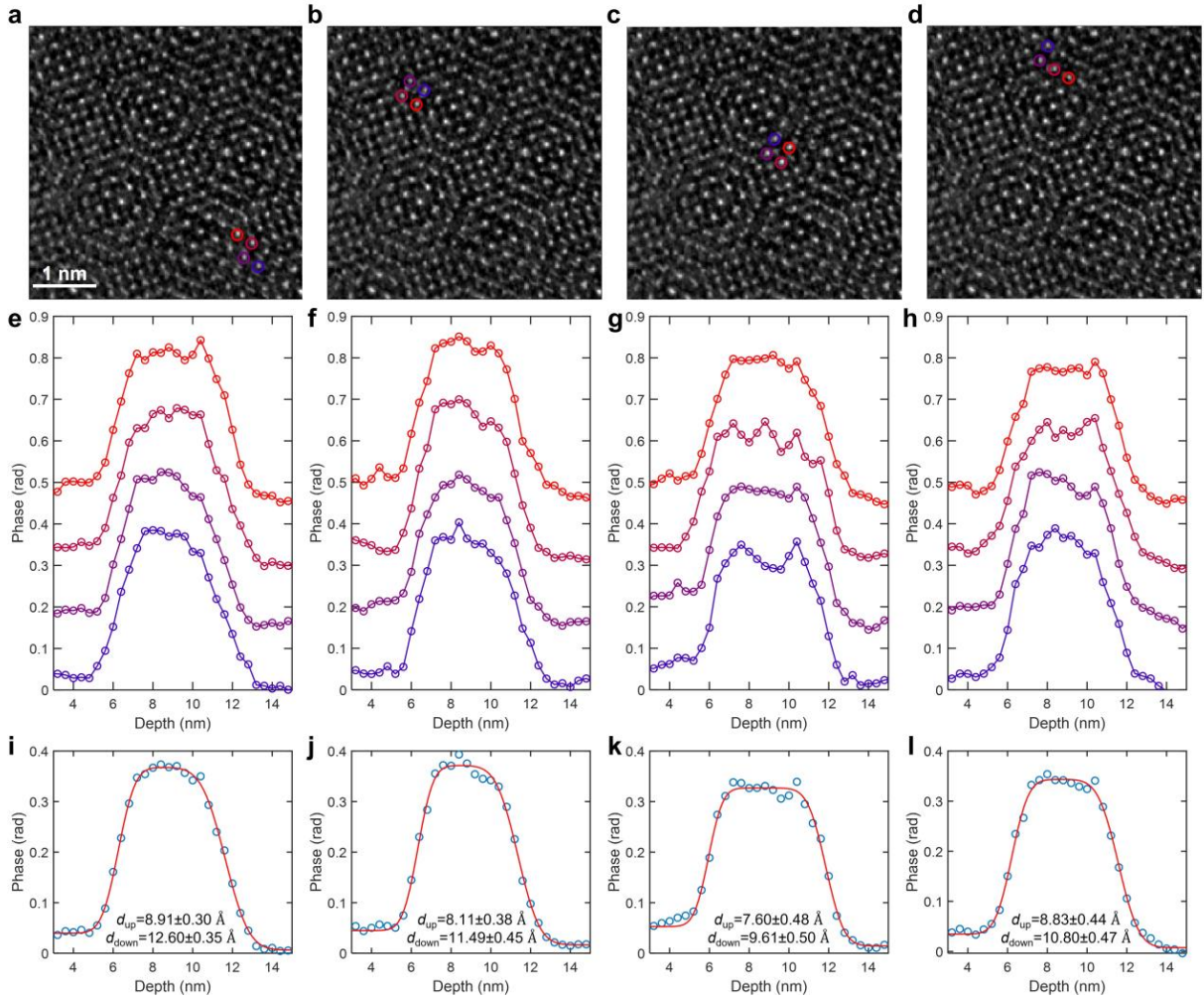

**Fig. S4| Atomic columns for determining the depth resolution with a maximum tilt angle of 2°. a-d**, Images of the Moiré pattern formed by the twisted double layer SrTiO$_3$ sample. The circles select several atomic columns from the bottom layer located near the edge of the Moiré pattern, which are well separated in the lateral dimension from other atoms in the top layer. **e-h**, Phase-depth curves of the selected atoms in (a)-(d) respectively, which are vertically shifted for clarity. **i-l**, Averaged phase-depth curve that is fitted by $y = A\,\mathrm{erf}[(x-\mu)/\sqrt{2}\sigma] + B$ on both the top and bottom surfaces, where **erf** denotes the error function defined by the integral of Gaussian function. Notably, the depth resolution from the top surface (i.e., from the interface) is slightly better than that from the bottom surface, which is chosen in the main text. Error bars are determined from curve fitting.

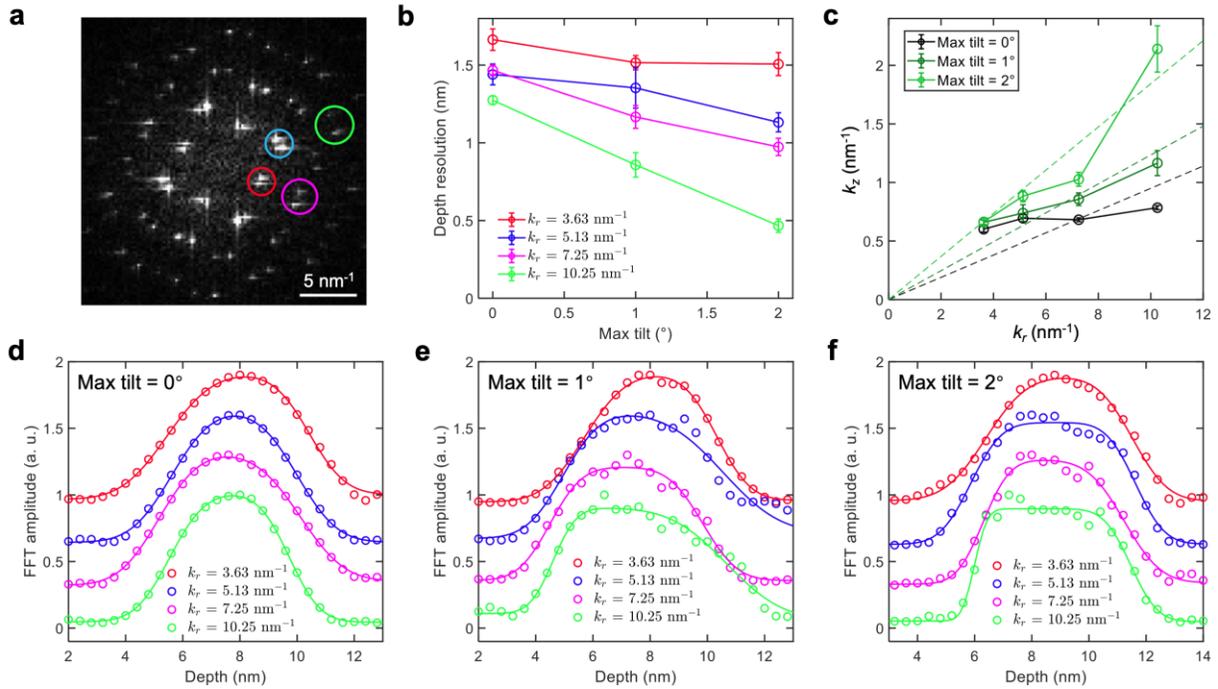

**Fig. S5| Detailed process for extracting the effective semi-angle of information transfer. a**, Fourier transformation of a slice with the presence of Moiré pattern, exhibiting distinct Bragg peaks. **b**, Depth resolution for each Bragg peak under different tilt conditions, which is extracted according to the procedure shown below in (d)-(f). **c**, Experimentally determined boundary of information transfer, where $k_z$ is defined as the reciprocal of depth resolution. **d**-**f**, Intensity of Bragg peaks versus depth, which is fitted by Error function to extract the depth resolution. The tilt angles used in the reconstructions are 1° (d), ±1° (e), ±1° and ±2° (f). Error bars are determined from curve fitting.

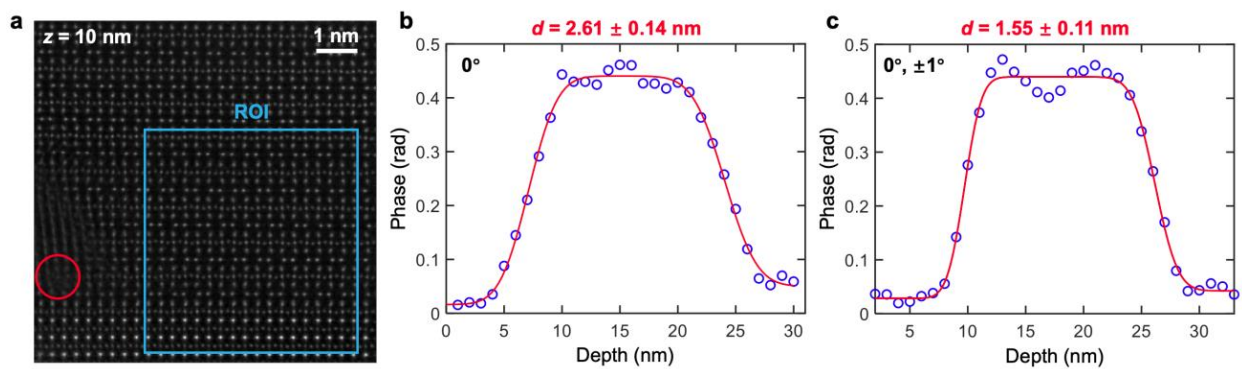

**Fig. S6| Overview of the reconstruction for (Pr,Ca)$_2$Co$_2$O$_5$/LaAlO$_3$ film. a**, Maximum field of view for the conventional MEP reconstruction. The red circle highlights the position of a dislocation, which is used later to align all datasets before the TCMEP reconstruction. The blue square represents the region of interest (ROI) in the main text, with a moderate distance away from the dislocation. **b-c**, Depth resolution for conventional MEP is ~3.0 nm (b), compared to ~1.7 nm (c) for TCMEP with a maximum tilt angle of 1°.

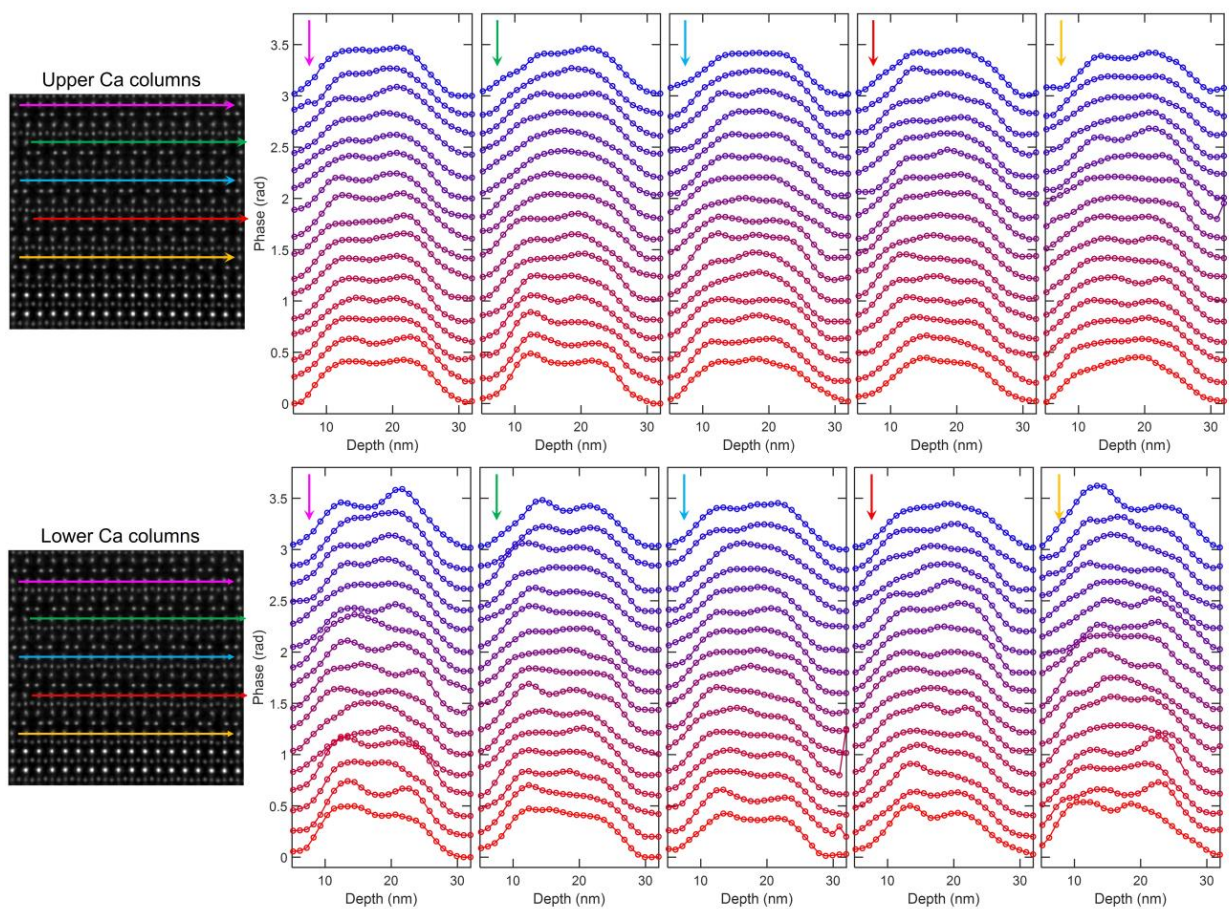

**Fig. S7| Phase depth curves from all Ca columns.** Left panels are phase images of a slice, where the phase depth curves of Ca atom columns along the arrows of different colors are plotted in the right panels respectively. The upper Ca columns exhibit smaller phase fluctuation than the lower one, indicating that Pr dopants primarily occupy the lower Ca columns within each unit cell.

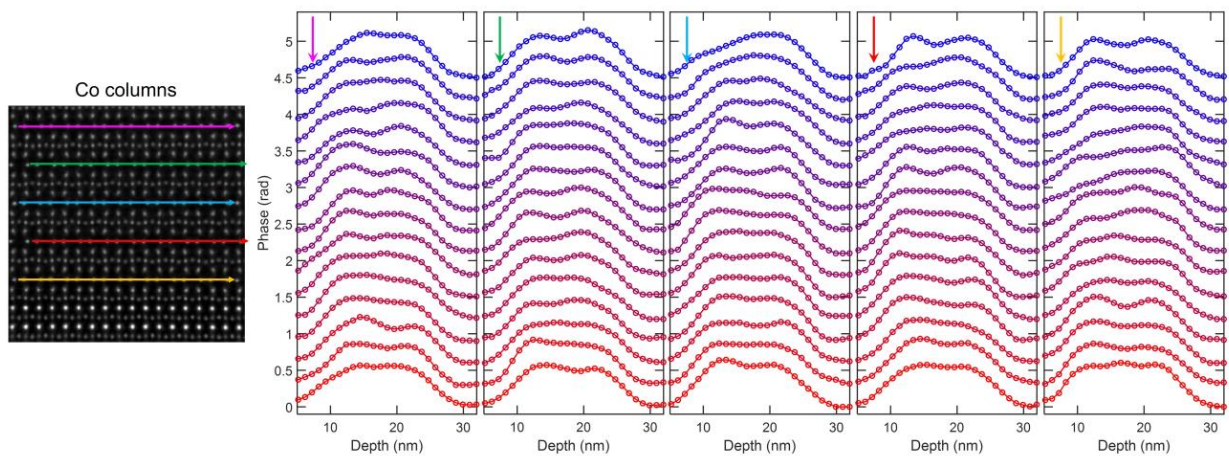

**Fig. S8| Phase depth curves from all Co columns.** Left panels are phase images of a slice, where the phase depth curves of Co atom columns along the arrows of different colors are plotted in the right panels respectively.

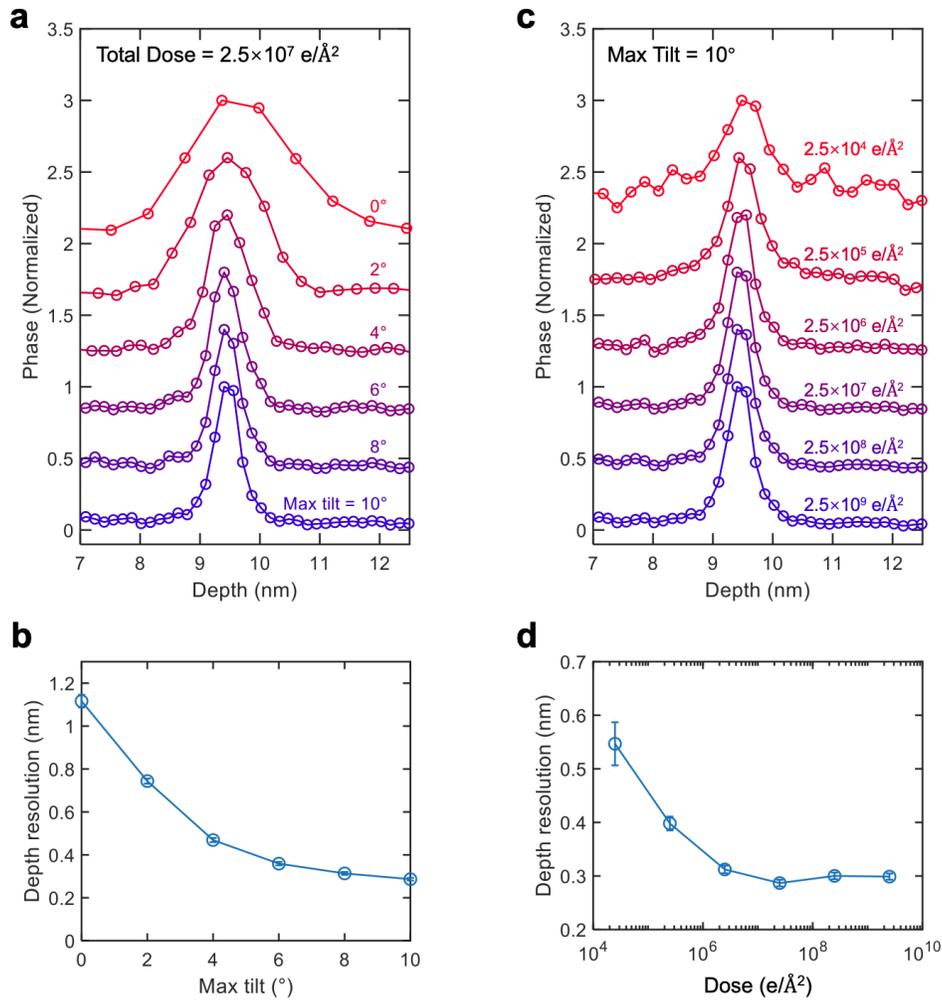

**Fig. S9| Simulation of TCMEP with larger tilt angles. a**, Phase-depth curves of the Sr dopant under various tilt conditions, while the total illumination dose remains at $2.5\times10^7$ $e/\text{Å}^2$. Curves are vertically shifted for clarity. **b**, Fitted depth resolution in (a) as a function of the maximum tilt angle. **c**, Phase-depth curves of the Sr dopant under various total illumination dose conditions, with a fixed maximum tilt angle of 10°. Curves are vertically shifted for clarity. **d**, Fitted depth resolution in **c** as a function of the total illumination dose. Error bars are determined from curve fitting.

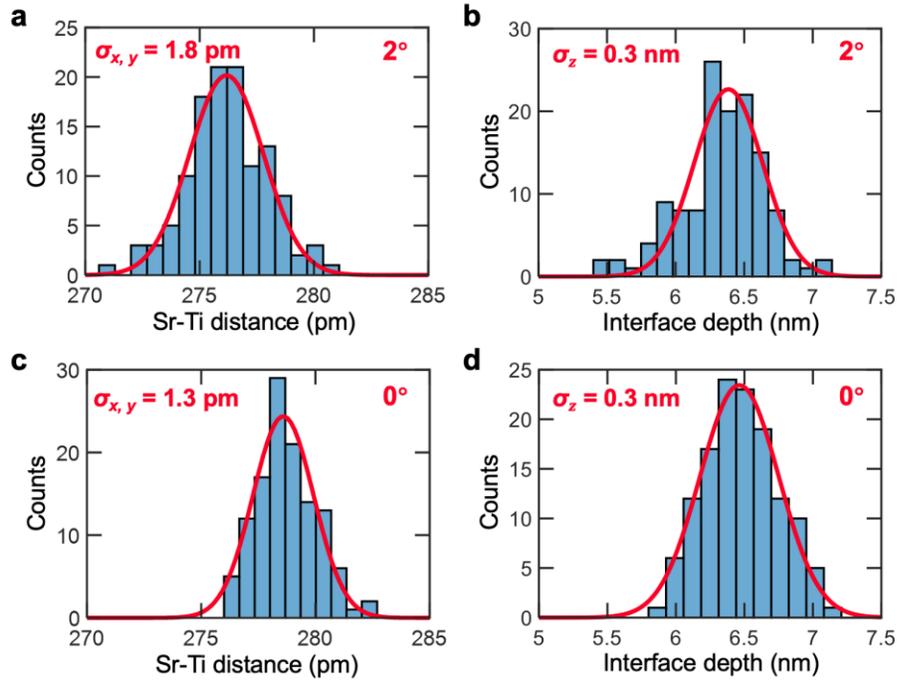

**Fig. S10| Precision of bond length measurements with TCMEP compared to MEP. a**, Histogram of the projected Sr-Ti distances from the bottom layer in the TCMEP reconstruction with a maximum tilt angle of 2°. The average distance is 276.2±1.8 pm, which agrees well with the STO lattice of 276.2 pm. **b**, Histogram of depth position of the interface in the TCMEP reconstruction with a maximum tilt angle of 2°, which is determined by fitting the phase-depth curves with error functions. The average depth is 6.3±0.3 nm. **c**, Histogram of the projected Sr-Ti distances with MEP. The average distance is 278.8±1.3 pm. **d**, Histogram of the depth position of the interface with MEP. The average depth is 6.5±0.3 nm.